\documentclass[oldversion]{aa} 
\usepackage{natbib}
\usepackage{graphicx}
\usepackage{txfonts}
\usepackage{calc}
\usepackage{color}
\usepackage{rotating}
\usepackage[OT2, T1]{fontenc}
\usepackage[USenglish]{babel} 
\usepackage{pdflscape}

\newcommand{\sunrise}{\textsc{Sunrise}}
\newcommand{\hinode}{\textsc{Hinode}}

\newcommand{\ftnt}[1]{#1\footnotemark\addtocounter{footnote}{-1}}

\begin{document}

\title{The potential of many-line inversions of photospheric spectropolarimetric data in the visible and near UV}
\titlerunning{Many-Line Inversions in the near UV} 

\author{T.~L. Riethm\"uller\inst{1}
   \and S.~K. Solanki\inst{1,2}
   }

\institute{Max-Planck-Institut f\"ur Sonnensystemforschung (MPS),
           Justus-von-Liebig-Weg 3, 37077 G\"ottingen, Germany
     \and
           School of Space Research, Kyung Hee University,
           Yongin, Gyeonggi, 446-701, Republic of Korea\\
           \email{riethmueller@mps.mpg.de}
          }

\date{Received; accepted}

\abstract
{
Our knowledge of the lower solar atmosphere is mainly obtained from spectropolarimetric observations, which are often carried out in the
red or infrared spectral range and almost always cover only a single or a few spectral lines. Here we compare the quality of Stokes inversions
of only a few spectral lines with many-line inversions. In connection with this, we also investigate the feasibility of spectropolarimetry in the short-wavelength range,
$3000\,\rm{\AA} - 4300\,\rm{\AA}$, where the line density but also the photon noise are considerably higher than in the red, so that many-line
inversions could be particularly attractive in that wavelength range. This is also timely because this wavelength range will be the focus of a new
spectropolarimeter in the third science flight of the balloon-borne solar observatory \sunrise{}. For an ensemble
of state-of-the-art magneto-hydrodynamical atmospheres we synthesize exemplarily spectral regions around 3140\,\AA{} (containing 371 identified
spectral lines), around 4080\,\AA{} (328 lines), and around 6302\,\AA{} (110 lines). The spectral coverage is chosen such that at a spectral
resolving power of 150000 the spectra can be recorded by a 2K $\times$ 2K detector. The synthetic Stokes profiles are degraded with a typical
photon noise and afterwards inverted. The atmospheric parameters of the inversion of noisy profiles are compared with the inversion of
noise-free spectra. We find that significantly more information can be obtained from many-line inversions than from a traditionally used
inversion of only a few spectral lines. We further find that information on the upper photosphere can be significantly more reliably obtained
at short wavelengths. In the mid and lower photosphere, the many-line approach at 4080\,\AA{} provides equally good results as the many-line approach at 6302\,\AA{}
for the magnetic field strength and the line-of-sight (LOS) velocity, while the temperature determination is even more precise by a factor of three.
We conclude from our results that many-line spectropolarimetry should be the preferred option in the future, and in particular at short wavelengths it
offers a high potential in solar physics.
}


\maketitle

\section{Introduction}

Gaining knowledge about the solar atmosphere requires the determination of its physical parameters such as temperature ($T$), magnetic field strength ($B$),
magnetic field inclination ($\gamma$), magnetic field azimuth ($\phi$), and line-of-sight velocity ($v_{\mathrm LOS}$). This is preferably achieved by
Stokes inversions \citep[see the recent review by][]{DelToro2016} of observational data recorded with spectropolarimeters, most of which can be classified
as either imaging spectropolarimeters (ISPs), or slit spectropolarimeters (SSPs).

To allow the investigation of dynamic processes, ISPs are mostly used at high cadences at the price of an only modest spectral resolution and spectral coverage.
The opposite is true for SSPs. It is only natural that most ISPs record only a few spectral lines because the wavelength interval is scanned by tuning a
variable filter, which usually takes time and significantly influences the cadence. Examples of ISPs are the instruments CRISP at the Swedish Solar Telescope
\citep{Scharmer2006}, IBIS at the Dunn Solar Telescope \citep{Cavallini2006}, IMaX onboard the balloon-borne \sunrise{} observatory \citep{MartinezPillet2011},
NFI onboard the \hinode{} satellite \citep{Tsuneta2008}, or HMI onboard the Solar Dynamics Observatory \citep{Scherrer2012}.

An SSP records the spectrum at a slit position quasi instantaneously. The number of spectral lines then only depends on the wavelength, the spectral resolution,
and the size of the detector, i.e. theoretically the recording of many spectral lines has long been possible. Numerous studies employing many spectral lines
have been carried out in the past, in particular using data gathered with a Fourier Transform Spectrometer \citep[FTS;][]{Brault1978}, e.g.,
\citet{Stenflo1983,Stenflo1984,Balthasar1984,Solanki1984,Solanki1985,Solanki1986}. However, more recently, observations of the solar photosphere at high spatial
resolution are almost entirely carried out with only a few spectral lines. Prominent examples are the instruments SP onboard the \hinode{} satellite \citep{Lites2013},
Trippel at the Swedish Solar Telescope \citep{Kiselman2011}, TIP2 at the Vacuum Tower Telescope \citep{Collados2007}, and GRIS at the GREGOR Solar Telescope
\citep{Collados2012}.

Reasons for this limitation to a few spectral lines are, firstly, that the photon flux available for ground-based observations
is decreased by the higher scattering of the blue part of sunlight in the Earth's atmosphere. Secondly, there is a focus on isolated (i.e. not blended)
spectral lines with a clear continuum, which are preferably found in the red and infrared spectral range. Only then is the use of classical
methods of estimating atmospheric parameters directly from the Stokes profiles possible \citep[see][for an overview of such techniques]{Solanki1993},
and also the usage of modern Stokes inversion codes is easier. Thirdly, rapid scanning of the evolving solar scene requires rapid read out, which can be achieved
by limiting the wavelength range. Fourthly, in the case of satellite-based instruments, limitations of the telemetry bandwidth can also provide a reason for a
reduced spectral coverage.

Because, obviously, a larger number of spectral lines potentially carries more information about the solar atmosphere and because of the improved technical possibilities,
increased attention has been paid to many-line approaches in recent times \citep[e.g.,][]{Beck2011,Rezaei2015,QuinteroNoda2017a,QuinteroNoda2017b}. Even observing many similar
spectral lines can be helpful in increasing the effective signal-to-noise (S/N) ratio without increasing the integration time, as has been demonstrated in cool-star
research \citep{Donati1997}.

By means of an ensemble of magneto-hydrodynamical (MHD) atmospheres, this paper considers quantitatively how well the many-line approach works and how large
its advantages (if any) are relative to traditionally used few-line approaches. By way of an example, we carry out our analysis for one spectral region each in the
red, in the violet, and in the near ultraviolet (NUV) and compare these. The examples are of general interest, but are chosen specifically to be also of relevance for
a new SSP, namely the \sunrise{} Ultraviolet Spectropolarimeter and Imager (SUSI). It is planned to operate SUSI onboard the balloon-borne solar observatory
\sunrise{} \citep{Barthol2011,Solanki2010,Solanki2017} during its third science flight at a float altitude above 30~km, so that access to the NUV is possible
and the scattering issue in the Earth's atmosphere does not exist. However, the analysis is general and its applications are not limited to a particular instrument.

In Sect.~\ref{Method} we present the method and introduce the employed MHD simulation as well as the spectral synthesis. Section~\ref{Results}
describes our results and in Sect.~\ref{Summary} we summarize the study and draw conclusions.

\section{Method}\label{Method}

We first consider the wavelength region $6280\,\rm{\AA} - 6323\,\rm{\AA}$. We chose this wavelength band because it contains the well-studied and frequently used
Fe\,{\sc i} line pair at 6302\,\AA{} that shows strong polarization signals and can be seen as a good example for Stokes inversions of spectropolarimetric data
in the red spectral range. The width of the wavelength range is chosen such that it can be critically sampled by a 2K $\times$ 2K pixels detector of an SSP at a spectral
resolving power of 150000 (i.e. each pixel covers $0.5 \lambda / 150000$), which we consider to be a reasonable compromise between a good spectral resolution and a broad
spectral coverage.

A three-dimensional (3D) MHD simulation of the photosphere and upper convection zone, which contains quiet-Sun regions (granulation) as well as strong-field features
(pores, bright points), was used for a spectral synthesis of the considered spectral region. After an estimation of the photon budget of a possible state-of-the-art
SSP, the synthetic Stokes profiles were degraded with photon noise. Among other tests, we then inverted the degraded synthetic spectra in order to retrieve
the atmospheric parameters.

The deviation of the inverted noise-contaminated atmospheres from the inverted noise-free atmospheres indicates how accurate the determination of temperature,
magnetic field vector, and LOS velocity in the photosphere is for the considered spectral region. The spectral range in the red was analyzed twice. Firstly,
for the many-line approach, i.e. for the full spectral region fitting on a 2K detector and containing 110 identified spectral lines, and secondly,
for a reduced spectral region ($6300.8\,\rm{\AA} - 6303.2\,\rm{\AA}$) that only covers the traditionally used Fe\,{\sc i} 6302\,\AA{} line pair.

\subsection{MHD Simulation}\label{Simulation}

Realistic simulations of the radiative and magneto-hydrodynamical processes in the photosphere and upper convection zone were carried out with the
3D non-ideal compressible MHD code MURaM \citep[The {\bf M}ax Planck Institute for Solar System Research / {\bf U}niversity of Chicago {\bf Ra}diation
{\bf M}agneto-hydrodynamics code; see][]{Voegler2005} that includes non-gray radiative energy transfer
under the assumption of local thermal equilibrium (LTE). Our simulation box covers 33.8\,Mm\,$\times$\,33.8\,Mm in its horizontal dimensions and
is 6.1\,Mm deep. The $\tau=1$ surface for the continuum at 5000\,\AA{} was on average reached about 700\,km below the upper boundary. The cell size of
the simulation box is 20.83\,km in the two horizontal directions and 16\,km in the vertical direction.

At the bottom boundary of the simulation box a free in- and outflow of matter was allowed under the constraint of total mass conservation, while
the top boundary was closed (zero vertical velocities). In the horizontal directions we used periodic boundary conditions.

\begin{figure}
\centering
\includegraphics[width=\linewidth]{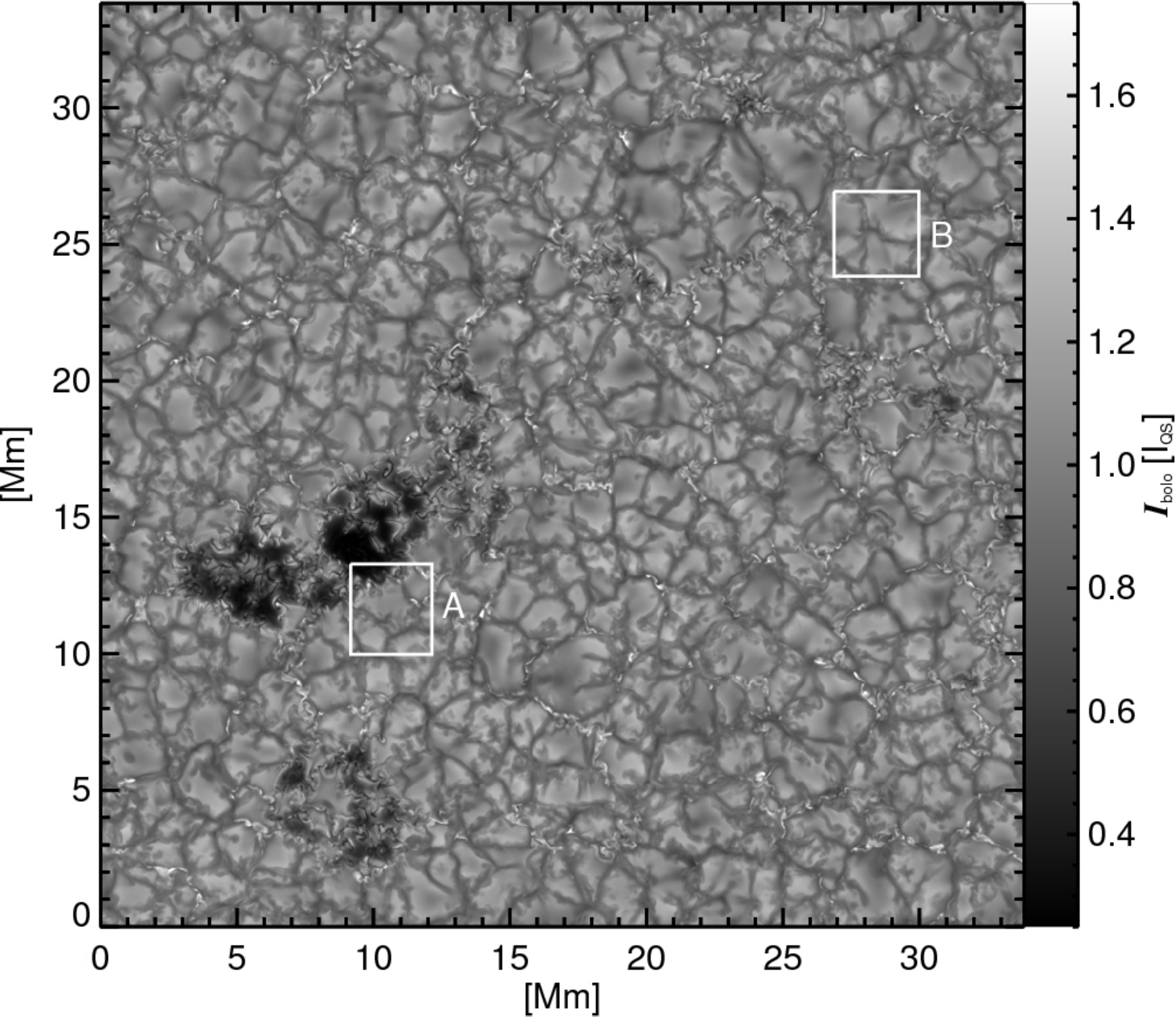}
\caption{Bolometric intensity image of the used MURaM simulation snapshot. The white boxes labeled with 'A' and 'B' indicate the regions of interest employed in this study.}
\label{Fig1}
\end{figure}

A spectropolarimetric observation of active region AR~11768 (with cosine of the heliocentric angle $\mu=0.93$), recorded on 2013 June 12, 23:39 UT
with the Imaging Magnetograph eXperiment \citep[IMaX;][]{MartinezPillet2011} onboard the \sunrise{} observatory
\citep{Solanki2010,Solanki2017,Barthol2011,Berkefeld2011,Gandorfer2011}, was inverted with the MHD-Assisted Stokes Inversion technique \citep[MASI;][]{Riethmueller2017b}.
This technique searches an archive of realistically degraded synthetic Stokes profiles for the best matches with the observed profiles. The best-fit MHD atmospheres
are used as the initial condition of our simulation. The simulation was then run for a further 54\,min of solar time to reach a statistically relaxed state.
This study uses a snapshot taken at this time. Fig.~\ref{Fig1} shows the map of bolometric intensities of the simulation snapshot. To limit the computational effort
needed for the current study we limit our analysis to the 3.0\,Mm\,$\times$\,3.3\,Mm wide region of interest indicated by the white box 'A' in Fig.~\ref{Fig1}, which,
in spite of its small size, contains part of a pore, some bright points, and seemingly normal granulation.

\subsection{Spectral Synthesis}\label{SpecSyn}

To obtain synthetic Stokes spectra from the MURaM simulation we used the inversion code
SPINOR \citep[{\bf S}tokes-{\bf P}rofiles-{\bf IN}version-{\bf O}-{\bf R}outines; see][]{Frutiger2000} in its forward calculation mode.
This code incorporates the STOPRO ({\bf STO}kes {\bf PRO}files) routines \citep{Solanki1987}, which compute synthetic Stokes profiles
of one or more spectral lines upon input of their atomic or molecular data and a model atmosphere. LTE conditions are assumed and the
Unno-Rachkovsky radiative transfer equations \citep{Rachkovsky1962} are solved. All spectral line syntheses in this paper were carried out
for the center of the solar disk ($\mu=1$).

We downloaded the atomic data of all spectral lines found in the atomic line databases of Kurucz \citep{Kurucz1995} and VALD
\citep[{\bf V}ienna {\bf A}tomic {\bf L}ine {\bf D}atabase; see][]{Ryabchikova2015} for the considered wavelength ranges. We performed a test-wise synthesis for
every single line contained in these databases for a hot standard atmosphere, the HSRA \citep{Chapman1979} extended to deeper layers representing the quiet Sun,
and also for a cold standard atmosphere, the Maltby-M \citep{Maltby1986} representing the core of an umbra. For these test computations we assigned a zero velocity
and magnetic field to the atmospheres. We ignored all spectral lines that could not be synthesized with SPINOR due to its limitations (e.g., SPINOR only supports
neutral and singly ionized atoms/molecules but no higher ionization states) and from the synthesizeable lines we further ignored insignificant ones in the sense that
we only considered spectral lines with a line depth larger than $10^{-2}$ in units of the continuum intensity in either of the two standard atmospheres. By applying
this procedure we identified 110 atomic lines for the $6280\,\rm{\AA} - 6323\,\rm{\AA}$ spectral region.
The solar abundance of all elements, including carbon, nitrogen, and oxygen, was taken from \citet{Grevesse1998}.

\begin{figure*}
\centering
\includegraphics[width=\linewidth]{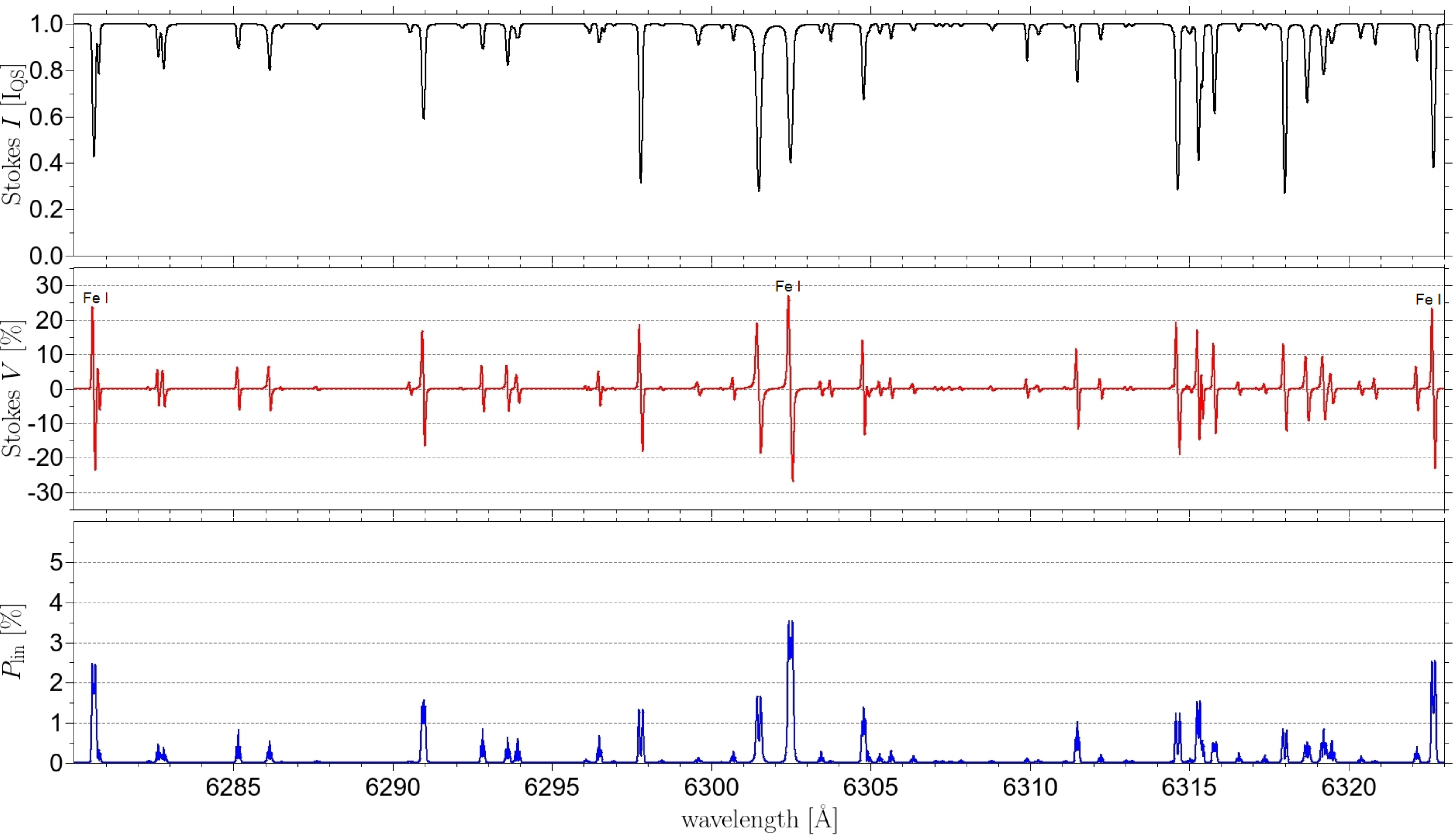}
\caption{Synthetic Stokes~$I$ (black line), Stokes~$V$ (red line), and linear polarization (blue line) profile at a spectral resolution of 150000
for a wavelength region around 6302\,\AA{} chosen to fit on a 2048 pixel wide detector if critically sampled in the wavelength domain. All spectra are normalized to the
mean quiet-Sun intensity, $I_{\rm{QS}}$. The horizontal lines in the lower two panels mark polarization levels of 10, 20, 30\,\% for Stokes~$V$ and
1, 2, 3, 4, 5\,\% for the linear polarization. Spectral lines providing $|V| > 20\,\%$ are labeled with the element forming the line.}
\label{Fig2}
\end{figure*}

The 110 spectral lines that where found synthesizeable by SPINOR and relevant for the solar photosphere were then synthesized again for an HSRA
temperature stratification and a zero velocity, but this time we assigned a magnetic field of 1\,kG strength and $30^{\circ}$ inclination. The magnetic field
properties of the atmosphere were taken to be constant with height and were selected such that significant polarization signals in Stokes~$Q$, $U$, and $V$
are reached. Also, after the radiative transfer computation, we spectrally degraded the Stokes profiles to a spectral resolving power of 150000.
The corresponding Stokes spectra are plotted in Fig.~\ref{Fig2}.
This spectral region contains the Fe\,{\sc i} 6301.5\,\AA{} and 6302.5\,\AA{} line pair, which is well studied, e.g. via the spectropolarimeter onboard the
\textsc{Hinode} satellite \citep{Tsuneta2008,Lites2013}. The amplitudes of the synthetic Stokes~$V$ signal of the two Fe\,{\sc i} lines are 18.9\,\% and 26.9\,\%, respectively,
while the amplitudes of the linear polarization ($P_{\mathrm{lin}} = \sqrt{Q^2+U^2}$) signal are 1.6\,\% and 3.5\,\%.

\subsection{Photon Noise}

Because the photon noise is calculated from the square root of the number of photo electrons, we compile the photon budget of a state-of-the-art
slit spectropolarimeter in Table~\ref{PhotonBudgetTable}. As an example, the used instrument parameters are based on the planned \sunrise{}/SUSI
instrument. We start our calculation with the solar spectral irradiance, $I_{s}$, which was measured by the third
{\bf AT}mospheric {\bf L}aboratory for {\bf A}pplication and {\bf S}cience mission \citep[ATLAS 3;][]{Thuillier2004,Thuillier2014}.
Since $I_{s}$ is the photon flux of the Sun integrated over the entire solar disk, we have to divide by the wavelength dependent limb-darkening
factor in order to get the photon flux at disk center (DC).

Parameters that can be influenced by the design of \sunrise{}/SUSI are the pixel size, the spectral resolution, the total transmission of the telescope
and the instrument, the quantum efficiency of the camera, the total integration time of the observation, and the polarimetric efficiency.

In the calculations an arc-second is fixed at 736.4\,km and the solar radius at 944.93421 arcsec, which corresponds to an observation time of
mid June, which is typical for a \sunrise{} flight in the Arctic. The primary mirror of the \sunrise{} telescope has a diameter of 100\,cm
and its central obscuration is 23.2\,cm across.

A cutting-edge SSP design includes a strictly synchronized, broadband slit-jaw camera that allows the determination of the point spread function (PSF).
The PSF can then be used to reconstruct the spectral data \citep{vanNoort2017} to correct for the residual wavefront aberrations. We expect an increase
in noise by the spectral restoration and take this into account in our photon budget by the noise amplification factor, $a$.

The calculation of the photon noise for the wavelength range centered on 6302\,\AA{} can be found in the third column of Table~\ref{PhotonBudgetTable}, while the
fourth and fifth column show the calculation for the wavelength bands around 4080\,\AA{} and 3140\,\AA{} that we will consider later in Sect.~\ref{CmpRedNuv}.

\begin{table*}
\begin{minipage}{\linewidth}
\renewcommand{\footnoterule}{\rule{1cm}{0cm} \vspace{-0cm}}
\setcounter{mpfootnote}{\value{footnote}+1}
\caption{Photon budget for the mean quiet-Sun intensity at disk center at the three wavelengths under study.}
\label{PhotonBudgetTable}
\begin{tabular}{l l l l l l} 
\hline
\noalign{\smallskip}
Quantity                                          & Unit             & 6302\,\AA{}                            & 4080\,\AA{}                            & 3140\,\AA{}                            & Source                                               \\
\hline                                           
\noalign{\smallskip}                             
Solar spectral irradiance ($I_{s}$)               & mW/m$^2$/nm      & 1657.8                                 & 1526.2                                 & 721.25                                 & ATLAS 3                                              \\
Limb darkening factor ($L$)                       &                  & 0.8238                                 & 0.7239                                 & 0.6739                                 & \citet{Cox1999}, Table 14.17                         \\
Solar flux at DC ($\Phi_1$)                       & mW/m$^2$/nm      & 2012.38                                & 2108.30                                & 1070.26                                & $I_{s}/L$                                            \\
Area of solar disk ($A_{\odot}$)                  & sr               & $6.59330\times 10^{-5}$                & $6.59330\times 10^{-5}$                & $6.59330\times 10^{-5}$                & $\pi\times(944.93421\times\pi/3600/180)^2$           \\
Solar flux at DC ($\Phi_2$)                       & J/s/cm$^2$/nm/sr & 3.05216                                & 3.19764                                & 1.62326                                & $\Phi_1/A_{\odot}$                                   \\
Photon energy ($E_{ph}$)                          & J/photon         & $3.1521\times 10^{-19}$                & $4.8688\times 10^{-19}$                & $6.3264\times 10^{-19}$                & $hc/\lambda$                                         \\
Diffraction limit                                 & km               & 117                                    & 76                                     & 58                                     & $1.22\lambda\times 3600\times 180\times 736.4/D/\pi$ \\
Pixel size ($P$)                                  & km/pixel         & 62.5                                   & 62.5                                   & 62.5                                   & $\approx$ critical sampling at 6302\,\AA{}           \\
Collecting area ($A_{col}$)                       & cm$^2$           & 7431.25                                & 7431.25                                & 7431.25                                & $(\pi/4)\times(100^2-23.2^2)$                        \\
Spectral resolving pow. ($R$)                     &                  & 150,000                                & 150,000                                & 150,000                                & $\lambda/\Delta\lambda$                              \\
Transmission ($Tr$)                               &                  & 0.2                                    & 0.2                                    & 0.2                                    & current \sunrise{}/SUSI design                       \\
Quantum efficiency ($QE$)                         &                  & 0.6                                    & 0.6                                    & 0.6                                    & own estimate                                         \\
Total integration time ($t_{int}$)                & s                & 4.8                                    & 4.8                                    & 4.8                                    & normal mode of \textsc{Hinode}/SP and SUSI           \\
Polar. eff. of Stokes~$I$ ($\epsilon_{I}$)        &                  & 1                                      & 1                                      & 1                                      & theoretical maximum                                  \\
Polar. eff. of $Q$,$U$,$V$ ($\epsilon_{QUV}$)     &                  & 0.57735                                & 0.57735                                & 0.57735                                & theoretical maximum ($\sqrt{1/3}$)                   \\
Energy per pixel and s ($E_{px}$)                 & J/pixel/s        & $161.339\times 10^{-13}$               & $109.432\times 10^{-13}$               & $42.753\times 10^{-13}$                & $\Phi_2 A_{col} \lambda (P/736.4\pi/180/3600)^2 /R$  \\
Photon rate ($\Phi_3$)                            & photons/s        & 51,183,920                             & 22,476,031                             & 6,757,967                              & $E_{px}/E_{ph}$                                      \\
\# of $e^-$ per integration ($\mathrm{N_{e^-}}$)  & electrons        & 29,481,941                             & 12,946,195                             & 3,892,589                              & $\Phi_3\times Tr\times QE\times t_{int}$             \\
Noise \ftnt{amplification} ($a$)                  &                  & 3                                      & 3                                      & 3                                      & \citet{MartinezPillet2011}                           \\
S/N of Stokes~$I$                                 &                  & 1810                                   & 1200                                   & 658                                    & $(\epsilon_{I}  \times \sqrt{\mathrm{N_{e^-}}}) / a$ \\
Noise of Stokes~$I$                               &                  & $5.5\times 10^{-4}$                    & $8.3\times 10^{-4}$                    & $1.5\times 10^{-3}$                    & $a / (\epsilon_{I}  \times \sqrt{\mathrm{N_{e^-}}})$ \\
Noise of Stokes~$Q$,$U$,$V$                       &                  & $9.6\times 10^{-4}$                    & $1.4\times 10^{-3}$                    & $2.6\times 10^{-3}$                    & $a / (\epsilon_{QUV}\times \sqrt{\mathrm{N_{e^-}}})$ \\
\hline                                                
\end{tabular}
\footnotetext{by image restoration of the spectra}
\setcounter{footnote}{\value{mpfootnote}}
\end{minipage}
\end{table*}

\subsection{Stokes Inversion}

After degrading the synthetic Stokes profiles with photon noise according to Table~\ref{PhotonBudgetTable}, the atmospheric
parameters were retrieved from Stokes inversions. For this step the SPINOR code was used in its inversion mode. We applied
an atmospheric model with five optical depth nodes at $\log\tau=-4, -2.5, -1.5, -0.8, 0$ for $T$, $B$, $\gamma$, $\phi$, and $v_{\mathrm LOS}$,
as well as a height-independent micro turbulence. All wavelengths were weighted equally. The synthetic Stokes spectra
that were calculated during the inversion were always convolved with a Gaussian profile that corresponds to a spectral resolving power of 150000.

The SPINOR inversion code was run twice in a row with 100 iterations each. While the first inversion run was carried out with uniform initial
conditions, a second run used the spatially smoothed inversion result of the first run as the initial guess atmosphere. This procedure removed
spatial discontinuities in the physical quantities that can occur if the inversion gets stuck in local minima of the merit function at individual
pixels or groups of pixels \citep[see also][]{Kahil2017,Solanki2017}.

\section{Results}\label{Results}

An ensemble of physically consistent MHD atmospheres of the region enclosed by the white box 'A' in Fig.~\ref{Fig1} covers important photospheric
features (quiet Sun, pores, bright points) and was used for the statistical analysis in this work.
All relevant spectral lines in the considered wavelength region were synthesized for this ensemble of atmospheres. The computed Stokes spectra were
spectrally degraded to a spectral resolving power of 150000 and the expected photon noise was added as given in Table~\ref{PhotonBudgetTable}.
The noisy Stokes spectra were inverted with the SPINOR code and the obtained atmospheric parameters were then compared with the inverted quantities
of the noise-free spectra.

For each atmospheric parameter and each optical depth node we subtract the maps of the noisy inversion from the noise-free inversion and calculate
the standard deviation, $\sigma$, over all pixels of such a difference image. In this way the inversion error purely induced by the noise can be expressed in the
form of a single number.

\subsection{Comparison of the many-line approach to the double-line approach in the red}\label{CompMultiDouble}

Table~\ref{SigmaTable6302v4BoxA} lists the inversion errors induced by photon noise for the many-line approach at 6302\,\AA{}, i.e. for the spectral region
$6280\,\rm{\AA} - 6323\,\rm{\AA}$ containing 110 considered spectral lines. We compare the many-line inversion with the same type of inversion
restricted to only two spectral lines ($6300.8\,\rm{\AA} - 6303.2\,\rm{\AA}$), i.e. the double-line approach, see Table~\ref{SigmaTable6302_2v4}.
The comparison strikingly shows the advantage of considering as many spectral lines as possible compared to a generally used single or double-line inversion.
$T$ as well as $B$ and $v_{\mathrm LOS}$ can be determined much better from the noisy Stokes profiles if all the lines are employed. Even for the magnetic
field inclination the many-line approach provides slightly better results. Since the computational effort of an inversion is influenced by the number of
the considered spectral lines, the many-line inversion consumes 22 times more computation time than the double-line approach, if all other inversion parameters
are kept the same.

\begin{table}
\caption{Noise-induced errors for the many-line inversion of region 'A' in Fig.~\ref{Fig1} at 6302\,\AA{} with five optical depth nodes.}
\label{SigmaTable6302v4BoxA}
\begin{tabular}{c c c c c} 
\hline
\noalign{\smallskip}
$\log{\tau}$ & T     & B   & $\gamma$ & $v_{\mathrm{LOS}}$ \\
             & [K]   & [G] & [deg]    & [m/s]              \\
\hline                               
\noalign{\smallskip}                 
-4           & 128   & 509 & 49.0     & 820                \\
-2.5         &  59.0 & 224 & 26.9     & 346                \\
-1.5         &  41.1 & 149 & 23.5     & 268                \\
-0.8         &  21.8 &  82 & 16.7     & 124                \\
  0          &   6.5 & 178 & 27.8     & 143                \\
\hline
\end{tabular}
\end{table}

\begin{table}
\caption{Same as Table~\ref{SigmaTable6302v4BoxA}, but for the double-line approach at 6302\,\AA{}.}
\label{SigmaTable6302_2v4}
\begin{tabular}{c c c c c} 
\hline
\noalign{\smallskip}
$\log{\tau}$ & T     & B   & $\gamma$ & $v_{\mathrm{LOS}}$ \\
             & [K]   & [G] & [deg]    & [m/s]              \\
\hline                               
\noalign{\smallskip}                 
-4           & 226   & 944 & 51.2     & 2170               \\
-2.5         &  95.1 & 275 & 30.8     &  621               \\
-1.5         & 118   & 231 & 26.7     &  484               \\
-0.8         & 130   & 175 & 19.2     &  357               \\
  0          &  43.3 & 308 & 29.5     &  587               \\
\hline
\end{tabular}
\end{table}

For both approaches, the temperature is best determined at $\log{\tau}=0$, as expected. There the temperature can be retrieved from noisy Stokes profiles by a factor of almost seven
more accurate if the entire spectral region is considered instead of only the one limited to the Fe\,{\sc i} line pair. $B$ and $v_{\mathrm LOS}$ are formed slightly higher in the
solar atmosphere, so that we find the smallest uncertainty at $\log{\tau}=-0.8$, where the accuracy of both quantities increases by a factor of roughly two by using all spectral lines
fitting on a 2K detector.

Generally, the $\sigma$ values in the upper photosphere (at $\log{\tau}=-4$) are larger than lower down in the atmosphere. Since this is particularly true for the double-line approach,
we consider if this is caused by the fact that spectra limited to only two spectral lines do not provide enough information for five optical depth nodes or, alternatively, if none
of the spectral lines from the chosen spectral range in the red provides information about the upper photosphere. To shed some light on this matter, we simplified our inversion model
for the double-line inversion to only four optical depth nodes at $\log\tau=-2.5, -1.5, -0.8, 0$ (see Table~\ref{SigmaTable6302_2v2}) and, even more restrictively, to only three
optical depth nodes at $\log\tau=-2.5, -1, 0$ (see Table~\ref{SigmaTable6302_2v3}), while all other inversion parameters remained unchanged.

\begin{table}
\caption{Same as Table~\ref{SigmaTable6302_2v4}, but for an inversion with only four optical depth nodes.}
\label{SigmaTable6302_2v2}
\begin{tabular}{c c c c c} 
\hline
\noalign{\smallskip}
$\log{\tau}$ & T     & B   & $\gamma$ & $v_{\mathrm{LOS}}$ \\
             & [K]   & [G] & [deg]    & [m/s]              \\
\hline                               
\noalign{\smallskip}                 
-2.5         & 35.0  & 340 & 40.8     & 516                \\
-1.5         & 39.0  & 229 & 23.4     & 442                \\
-0.8         & 51.3  & 172 & 20.4     & 318                \\
  0          & 18.1  & 334 & 32.6     & 580                \\
\hline
\end{tabular}
\end{table}

\begin{table}
\caption{Same as Table~\ref{SigmaTable6302_2v4}, but for an inversion with only three optical depth nodes.}
\label{SigmaTable6302_2v3}
\begin{tabular}{c c c c c} 
\hline
\noalign{\smallskip}
$\log{\tau}$ & T     & B   & $\gamma$ & $v_{\mathrm{LOS}}$ \\
             & [K]   & [G] & [deg]    & [m/s]              \\
\hline                               
\noalign{\smallskip}                 
-2.5         & 22.3  & 419 & 43.0     & 793                \\
-1           & 29.2  &  82 & 11.7     & 156                \\
 0           &  9.8  & 249 & 17.7     & 461                \\
\hline
\end{tabular}
\end{table}

The reduction of the number of nodes leads to significantly more accurate $T$ values at all considered optical depths (at the price of a reduced knowledge of the height
dependence), while we find an ambiguous picture for $B$. At $\log{\tau}=-1$, the standard deviation of $B$ is only half of the corresponding value for the five-node inversion
($\log{\tau}=-0.8$) at the expense of worse results for $B$ somewhat higher up at $\log{\tau}=-2.5$. For $v_{\mathrm LOS}$ we find a slight improvement when reducing
the number of nodes from five to four, but for the three-node inversion the situation is similar to the one for $B$.

To limit the computational effort, we so far considered only a relatively small region of interest (white box labeled with 'A' in Fig.~\ref{Fig1}). We now test the statistical significance
of our analysis by repeating the many-line inversion at 6302\,\AA{} for another region of almost the same size (box 'B' in Fig.~\ref{Fig1}). The standard deviations are listed in
Table~\ref{SigmaTable6302v4BoxB} and show almost identical values for the temperature and also for the magnetic field inclination. The values for $B$ and $v_{\mathrm{LOS}}$ seem to be better
determined for region 'B', probably because this is a pure quiet-Sun region that does not contain any dark strong-field features as we have in region 'A'. The S/N ratio of such dark features
can be much worse than for quite-Sun regions which reduces the accuracy of the retrieved atmospheric quantities.

\begin{table}
\caption{Same as Table~\ref{SigmaTable6302v4BoxA}, but for region 'B' in Fig.~\ref{Fig1}.}
\label{SigmaTable6302v4BoxB}
\begin{tabular}{c c c c c} 
\hline
\noalign{\smallskip}
$\log{\tau}$ & T     & B   & $\gamma$ & $v_{\mathrm{LOS}}$ \\    
             & [K]   & [G] & [deg]    & [m/s]              \\    
\hline                                                           %
\noalign{\smallskip}                                             %
-4           & 134   & 332 & 52.8     & 546                \\    
-2.5         &  62.3 & 100 & 29.5     & 199                \\    
-1.5         &  43.7 &  87 & 23.8     & 166                \\    
-0.8         &  21.7 &  55 & 14.5     &  84                \\    
  0          &   6.1 & 156 & 28.3     & 124                \\    
\hline
\end{tabular}
\end{table}

\subsection{Comparison between the red and (ultra)violet spectral bands}\label{CmpRedNuv}

The Planck function falls steeply towards short wavelengths, so that spectropolarimetric measurements in the NUV always suffer from the problem
of high photon noise relative to the signal. At the same time the spectral line density increases strongly in the NUV. We showed in Sect.~\ref{CompMultiDouble}
that the quality of the inversion results in the red spectral range improves significantly if we consider 110 spectral lines instead of only two.
Based on that result, we now investigate to what extent the loss of information caused by high photon noise can be compensated by the gain in information due to an increased
number of the observed spectral lines. Other effects that also can influence the quality of spectropolarimetric observations, e.g., stray light, wavefront aberrations,
or the wavelength dependence of the diffraction limit are again not considered.

To identify interesting spectral regions in the violet and NUV spectral range, we once more calculated synthetic Stokes profiles for the 1\,kG HSRA atmosphere described in
Sect.~\ref{SpecSyn}. We did this for the spectral range $3000\,\rm{\AA} - 4300\,\rm{\AA}$ that will be covered by the planned slit spectropolarimeter, SUSI,
which is to partake on the third science flight of \sunrise{}. In addition to the atomic lines of the Kurucz and VALD databases,
we also considered some important molecular lines that were identified in previous studies, in particular 139 lines formed by the OH molecule in the spectral region around
3120\,\AA{} and 233 lines of the CN molecule around 3880\,\AA{} \citep[for details see][]{Riethmueller2014} as well as 241 CH lines in the so-called G-band, a spectral
region around 4300\,\AA{} \citep{Shelyag2004,Riethmueller2017a}.

\begin{figure*}
\centering
\includegraphics[width=\linewidth]{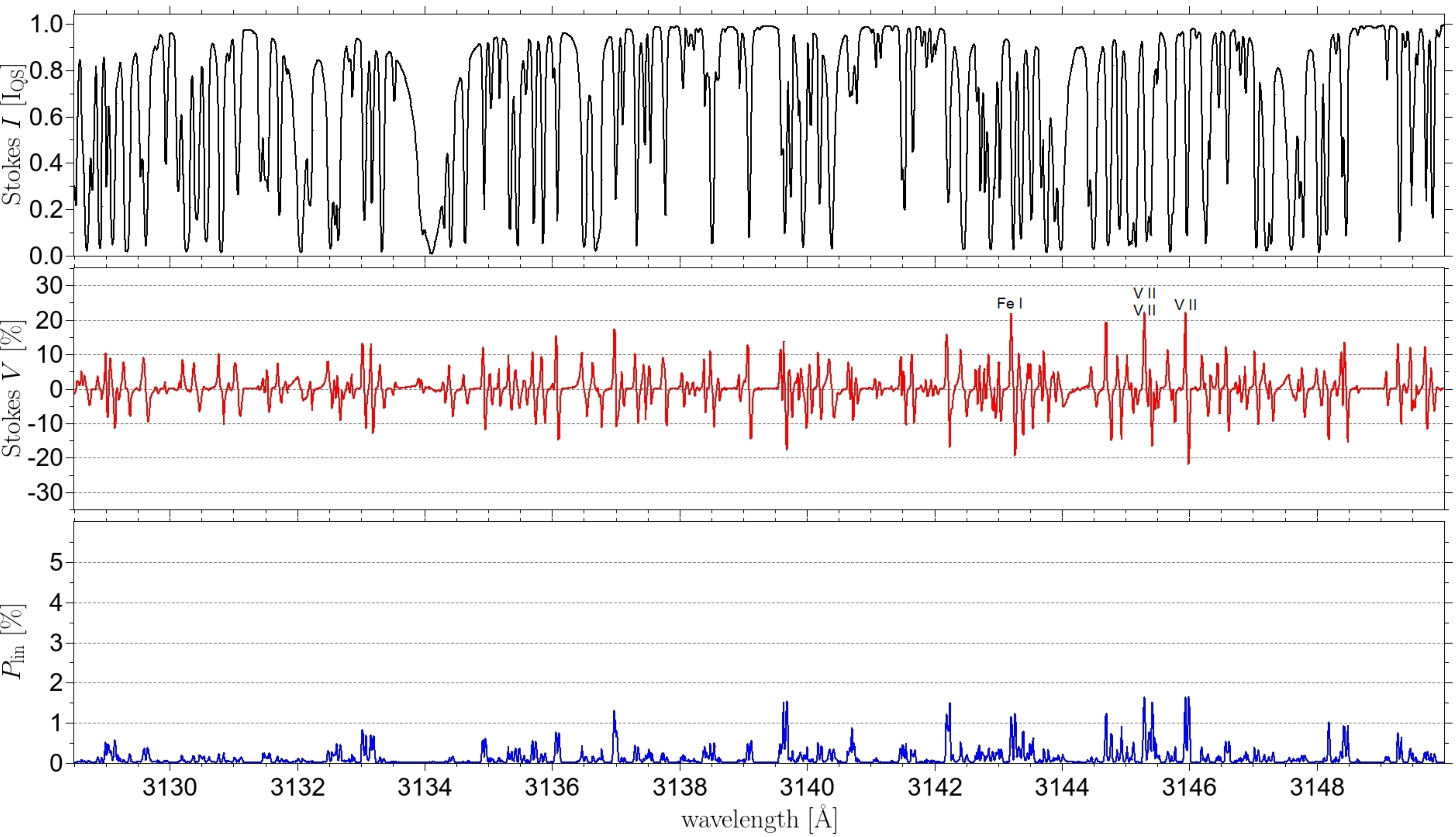}
\caption{Same as Fig.~\ref{Fig2}, but for a spectral region around 3140\,\AA{}.}
\label{Fig3}
\end{figure*}

We searched the wavelength range $3000\,\rm{\AA} - 4300\,\rm{\AA}$ for spectral regions fitting on a 2K detector at a critical sampling in the wavelength dimension
and exhibiting as many strongly polarized spectral lines as possible.
We found several such spectral regions, but limit our study to two of them due to the large computational effort.
In the following we focus on the spectral region $3128.5\,\rm{\AA} - 3150\,\rm{\AA}$, which is relatively close to the lower wavelength limit of SUSI and hence
can potentially provide the highest spatial resolution. At the same time, however, we expect the largest problems with photon noise in this spectral region.
Fig.~\ref{Fig3} shows the synthetic Stokes profiles of this spectral region. 42 of the 371 synthesized spectral lines show a Stokes~$V$ signal larger than 10\,\%.
Three of them provide a Stokes~$V$ signal even larger than 20\,\% and are listed with their Land\'e factors in Table~\ref{StrongVLineTable} and, in addition,
they are labeled in the middle panel of Fig.\ref{Fig3}. A $P_{\mathrm{lin}}$ signal higher than 1\,\% is reached by 9 spectral lines.

\begin{figure*}
\centering
\includegraphics[width=\linewidth]{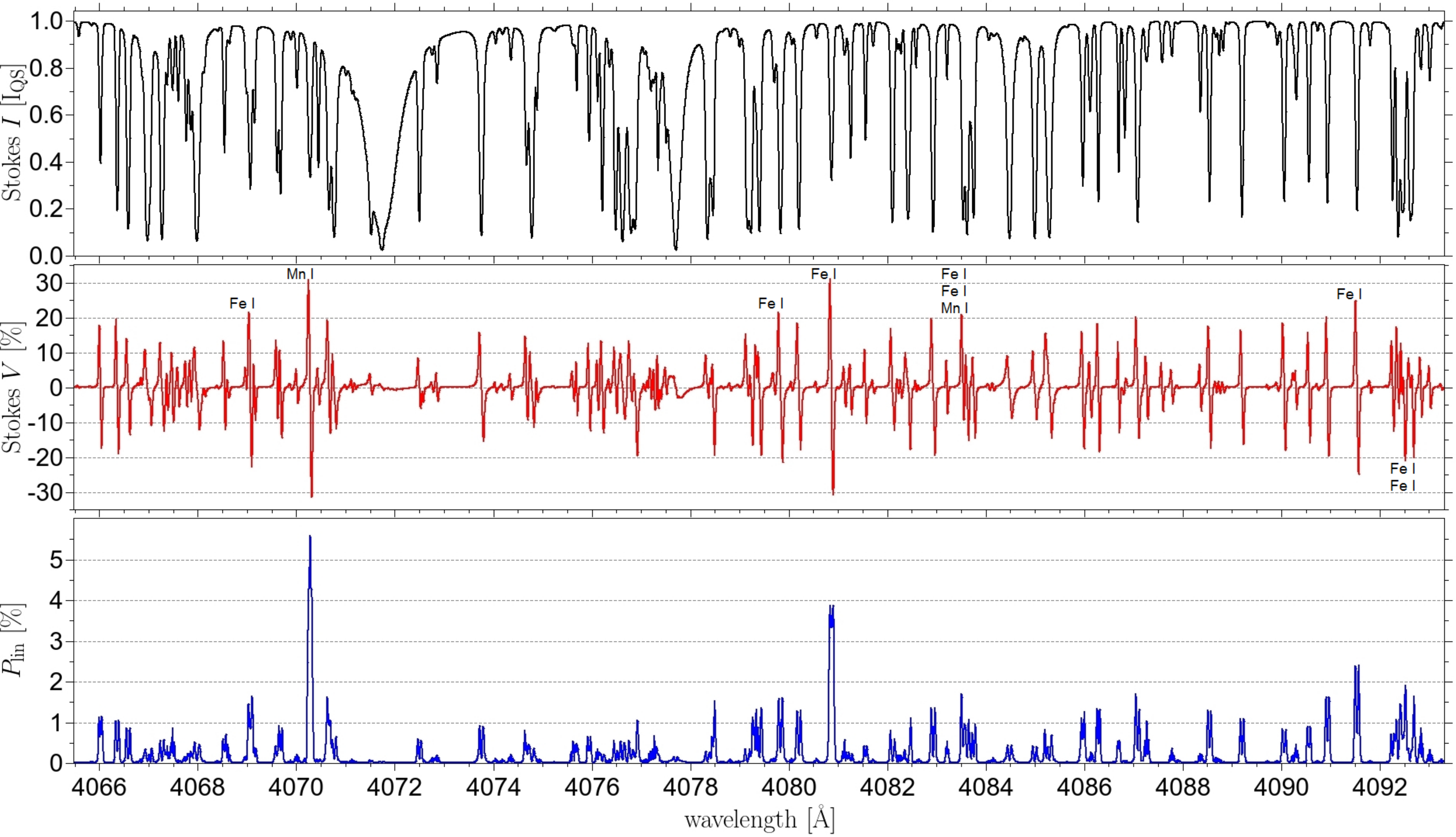}
\caption{Same as Fig.~\ref{Fig2}, but for a spectral region around 4080\,\AA{}.}
\label{Fig4}
\end{figure*}

The second spectral region we consider is $4065.49\,\rm{\AA} - 4093.31\,\rm{\AA}$ because between $3000\,\rm{\AA}$ and $4300\,\rm{\AA}$ it hosts
the largest number of strongly polarized spectral lines. Since this region is not too far from the upper limit of the SUSI wavelength range, it has the advantage
of a much higher photon flux, but at the price of a somewhat worse diffraction limit. We synthesized 328 spectral lines in this part of the spectrum and plotted the
Stokes profiles in Fig.~\ref{Fig4}. We find a similar number of lines exceeding the 10\,\% level in Stokes~$V$ as around $3140\,\rm{\AA}$, but this time seven of them
show a Stokes~$V$ signal larger than 20\,\%, a much higher fraction than for the 3140\,\AA{} region. In addition, two lines display a $P_{\mathrm{lin}}$ signal larger
than 3\,\%, namely the Mn\,{\sc i} 4070.279\,\AA{} line and the Fe\,{\sc i} 4080.875\,\AA{} line (see Table~\ref{StrongVLineTable} for their Land\'e factors).

\begin{table*}
\caption{Central wavelengths and Land\'e factors, $g$, of considered spectral lines providing a Stokes~$V$ signal higher than 20\,\%. Lines that overlap each other (blends) are listed in a row.}
\label{StrongVLineTable}
\begin{tabular}{l l l l l l l l l} 
\hline                                                   
\noalign{\smallskip}                                     
3143.2425\,\AA & Fe\,{\sc i} & $g=2.1250$  &                &             &             &                &             &            \\
3145.3260\,\AA & V\,{\sc ii} & $g=2.1667$; & 3145.3510\,\AA & V\,{\sc ii} & $g=1.9000$  &                &             &            \\
3145.9690\,\AA & V\,{\sc ii} & $g=2.0000$  &                &             &             &                &             &            \\
\noalign{\smallskip}                                     
\hline                                                   
\noalign{\smallskip}                                     
4069.0650\,\AA & Fe\,{\sc i} & $g=2.2500$  &                &             &             &                &             &            \\
4070.2790\,\AA & Mn\,{\sc i} & $g=3.3333$  &                &             &             &                &             &            \\
4079.8384\,\AA & Fe\,{\sc i} & $g=2.0000$  &                &             &             &                &             &            \\
4080.8749\,\AA & Fe\,{\sc i} & $g=3.0000$  &                &             &             &                &             &            \\
4083.5421\,\AA & Fe\,{\sc i} & $g=1.5000$; & 4083.5489\,\AA & Fe\,{\sc i} & $g=1.5729$; & 4083.6290\,\AA & Mn\,{\sc i} & $g=1.5000$ \\
4091.5527\,\AA & Fe\,{\sc i} & $g=2.0000$  &                &             &             &                &             &            \\
4092.4562\,\AA & Fe\,{\sc i} & $g=1.7500$; & 4092.5068\,\AA & Fe\,{\sc i} & $g=1.8333$  &                &             &            \\
\noalign{\smallskip}                                     
\hline                                                   
\noalign{\smallskip}                                     
6280.6163\,\AA & Fe\,{\sc i} & $g=1.4500$  &                &             &             &                &             &            \\
6302.4936\,\AA & Fe\,{\sc i} & $g=2.5000$  &                &             &             &                &             &            \\
6322.6873\,\AA & Fe\,{\sc i} & $g=1.5000$  &                &             &             &                &             &            \\
\noalign{\smallskip}                                     
\hline
\end{tabular}
\end{table*}

In Table~\ref{NumberTable} we contrast the number of strongly polarized spectral lines for the three spectral regions under study. Surprisingly,
the 4080\,\AA{} region contains more lines than the 6302\,\AA{} region with polarization signals above a certain level, while the 3140\,\AA{} region is more or less
comparable to the one at 6302\,\AA{}.

\begin{table}
\caption{Number of spectral lines found in the three considered wavelength bands that have polarization signals higher than the listed levels.}
\label{NumberTable}
\begin{tabular}{c c c c} 
\hline
\noalign{\smallskip}
Polarization               & 3140\,\AA{} & 4080\,\AA{}  & 6302\,\AA{}    \\
\hline                                                   
\noalign{\smallskip}                                     
$|V|$ > 30\,\%             & 0           & 2            & 0              \\
$|V|$ > 20\,\%             & 3           & 7            & 3              \\
$|V|$ > 10\,\%             & 42          & 48           & 11             \\
\hline                                                   
$P_{\mathrm{lin}}$ > 5\,\% & 0           & 1            & 0              \\
$P_{\mathrm{lin}}$ > 4\,\% & 0           & 1            & 0              \\
$P_{\mathrm{lin}}$ > 3\,\% & 0           & 2            & 1              \\
$P_{\mathrm{lin}}$ > 2\,\% & 0           & 3            & 3              \\
$P_{\mathrm{lin}}$ > 1\,\% & 9           & 22           & 8              \\
\hline
\end{tabular}
\end{table}

The basic aim of this part of the study is to compare the results of inversions at the three wavelengths. To do this comparison in
a fair manner, we model these wavelength bands as if they were observed by a single instrument. As an example we use the parameters of \sunrise{}/SUSI
(which, however, will not observe at 6300\AA) and take into account the different noise levels at the three wavelengths etc. We do not, however,
scale the pixel size or spatial resolution with wavelength, keeping these two the same, so as to be able to compare the capabilities on an equal
footing. I.e., the pixel size for all three analyzed spectral regions is fixed at 62.5\,km (see Table~\ref{PhotonBudgetTable}), which roughly
corresponds to a critical sampling at 6302\,\AA{}. This means, that the SUSI data, for which a critical sampling at 3000\,\AA{} is planned,
are assumed to be binned.

The wavelength bands around 3140\,\AA{} and 4080\,\AA{} were identically treated as in the red, i.e. the ensemble of MHD atmospheres of the region
enclosed by the white box 'A' in Fig.~\ref{Fig1} was synthesized, spectrally degraded, contaminated with photon noise, and inverted. Differences
between the inversion results of noisy and noise-free profiles were used to calculate standard deviations. The hope, that the increased
number of considered spectral lines compensates for the larger noise, came from the fact that the photon noise at 3140\,\AA{} is higher by a factor
of 2.8 than in the red at 6302\,\AA{} (Table~\ref{PhotonBudgetTable}), while the number of spectral lines identified for the 3140\,\AA{} region, however,
is larger by a factor of $371/110=3.4$.

The standard deviations of the many-line inversion in the ultraviolet spectral range around 3140\,\AA{} with five optical depth nodes are shown in Table~\ref{SigmaTable3140v4}.
The results of this inversion are competitive to the double-line inversion results in the red (Table~\ref{SigmaTable6302_2v4}) because they exhibit similarly good results for
$B$ and $\gamma$, and an outcome for $T$ and $v_{\mathrm LOS}$ that is better by factors of two to eight. Compared to a many-line inversion in the red (Table~\ref{SigmaTable6302v4BoxA}),
$T$ and $v_{\mathrm LOS}$ can be determined more accurately at 3140\,\AA{}, but for $B$ the UV inversion is only more accurate at $\log{\tau}=-4$.

\begin{table}
\caption{Same as Table~\ref{SigmaTable6302v4BoxA}, but for the many-line approach at 3140\,\AA{}.}
\label{SigmaTable3140v4}
\begin{tabular}{c c c c c} 
\hline
\noalign{\smallskip}
$\log{\tau}$ & T    & B   & $\gamma$ & $v_{\mathrm{LOS}}$ \\
             & [K]  & [G] & [deg]    & [m/s]              \\
\hline                               
\noalign{\smallskip}                 
-4           & 65.5 & 367 & 45.5     & 276                \\
-2.5         & 34.4 & 368 & 27.9     & 271                \\
-1.5         & 21.1 & 334 & 25.4     & 245                \\
-0.8         & 14.1 & 159 & 18.4     & 117                \\
  0          &  6.7 & 307 & 26.1     & 106                \\
\hline
\end{tabular}
\end{table}

The most reliable determination of the atmospheric parameters is provided by a many-line inversion at 4080\,\AA{} (Table~\ref{SigmaTable4080v4}). All quantities show much
lower standard deviations than for a double-line inversion in the red (Table~\ref{SigmaTable6302_2v4}). Even compared to a many-line inversion in the red (Table~\ref{SigmaTable6302v4BoxA}),
we generally find considerably lower standard deviations at $\log{\tau}=-4$. The other four optical depth nodes provide clearly more accurate $T$ results at 4080\,\AA{}, while
the determination of $B$ and $v_{\mathrm LOS}$ is more or less equally good as in the red.

\begin{table}
\caption{Same as Table~\ref{SigmaTable6302v4BoxA}, but for the many-line approach at 4080\,\AA{}.}
\label{SigmaTable4080v4}
\begin{tabular}{c c c c c} 
\hline
\noalign{\smallskip}
$\log{\tau}$ & T    & B   & $\gamma$ & $v_{\mathrm{LOS}}$ \\
             & [K]  & [G] & [deg]    & [m/s]              \\
\hline                               
\noalign{\smallskip}                 
-4           & 40.7 & 428 & 44.6     & 342                \\
-2.5         & 20.3 & 237 & 26.6     & 231                \\
-1.5         & 15.3 & 173 & 22.9     & 281                \\
-0.8         &  7.9 &  90 & 15.8     & 119                \\
  0          &  2.4 & 175 & 23.8     & 193                \\
\hline
\end{tabular}
\end{table}

\begin{figure*}
\centering
\includegraphics[width=\linewidth]{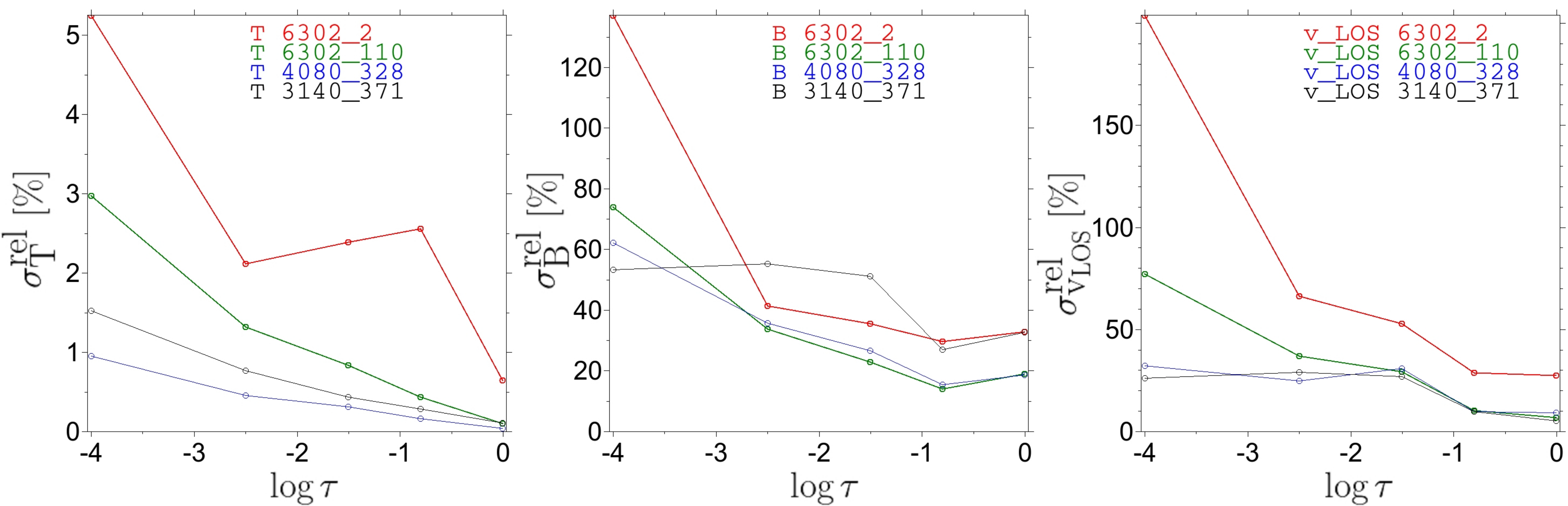}
\caption{Relative inversion errors as function of the optical depth for the the many-line approaches at 3140\,\AA{} (black lines),
at 4080\,\AA{} (blue lines), and at 6302\,\AA{} (green lines) as well as for the double-line approach at 6302\,\AA{} (red lines).
The errors of the three quantities temperature, magnetic field strength and LOS velocity are shown in the left, middle, and right panel,
respectively.}
\label{Fig5}
\end{figure*}

For an overview of the inversion results, which compares the various approaches and optical depths, we divided the standard deviations by the spatial mean
of the unsigned quantities to get relative numbers. For the inversions with five optical depth nodes, the relative standard deviations of
$T$, $B$, and $v_{\mathrm LOS}$ are plotted in Fig.~\ref{Fig5} and show the superiority of the many-line approach (green line) over the double-line approach
(red line) in the red. In the upper photosphere at $\log\tau=-4$, the gain in information due to the consideration of many spectral lines is particularly
strong, but also observations at shorter wavelengths improve the inversion results in the upper photosphere, at least for $T$ and $v_{\mathrm LOS}$.
If we consider all three atmospheric quantities at all five optical depths nodes, Fig.~\ref{Fig5} demonstrates the outstanding results of the many-line approach
at 4080\,\AA{} compared to the other wavelengths and even more compared with the double-line approach.

For the many-line inversions in the red, violet, and NUV we used a cluster of 96 cores (AMD Opteron 6176). To invert the Stokes profiles of
a single ray in a single inversion run with 100 iterations the mean computational time on this cluster was 85\,s at 3140\,\AA{}, 20\,s at 4080\,\AA{},
and 11\,s at 6302\,\AA{}.

\subsection{Scatterplots}

\begin{figure*}
\centering
\includegraphics[width=0.79\linewidth]{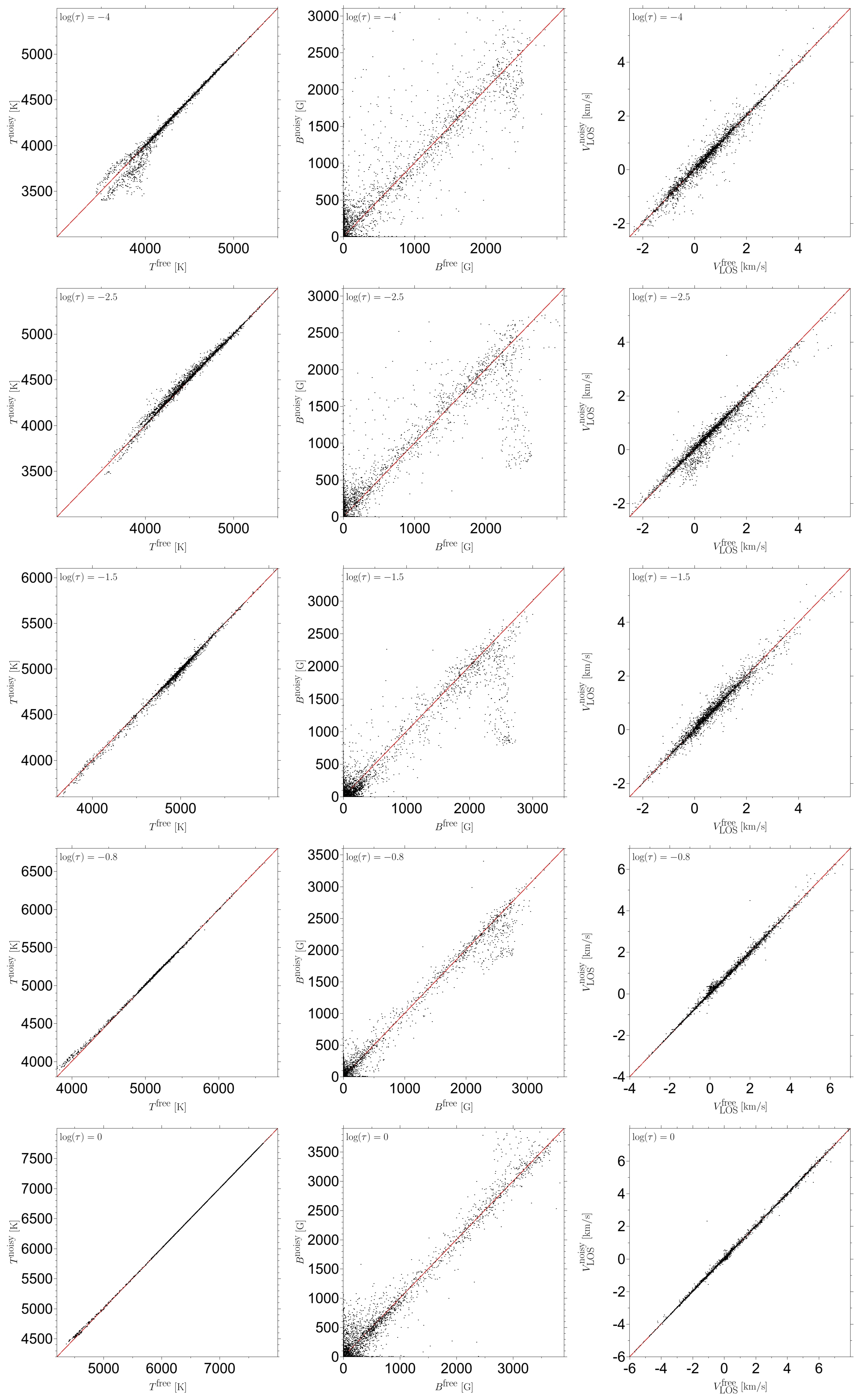}
\caption{Scatter plots (black dots) of the temperature, magnetic field strength, and LOS velocity (from left to right) retrieved from an inversion
of synthetic Stokes profiles degraded with realistic photon noise versus the inversion results from noise-free profiles at the optical depth nodes used
during the inversion for the spectral region $3128.5\,\rm{\AA} - 3150\,\rm{\AA}$ that contains 371 relevant spectral lines. The ideal relation for a perfect
fit is indicated by the red line. From top to bottom the panels refer to different optical depth nodes (marked in the figures).}
\label{Fig6}
\end{figure*}

Since the standard deviation of a difference image is only a single number that summarizes the deviation of the inversion of noisy spectra from the inversion of
noise-free profiles, we now show scatterplots of $T$, $B$, and $v_{\mathrm LOS}$ at the optical depths chosen for the inversion. We do this for two examples of particular
interest. First for the many-line approach at 3140\,\AA{}, which is the shortest wavelength under study. The five optical depth nodes at $\log\tau=-4, -2.5, -1.5, -0.8, 0$
are shown in Fig.~\ref{Fig6} from top to bottom (see text labels). The scatter of $T$ and $v_{\mathrm LOS}$ is relatively low and decreases as the optical depth increases.
At optical depth unity, almost no deviations can be found between the results of the noisy and noise-free inversion for $T$ and we see only a small scatter for
$v_{\mathrm LOS}$. The photon noise causes difficulties, in particular to the determination of $B$, where the largest deviations from the noise-free inversion
are found in the upper atmospheric layers and at higher field strengths. Values greater than $B>2$\,kG are only reached in pores, where the continuum intensity at 3140\,\AA{}
can drop to 3.3\,\% of the mean quiet-Sun value, which further reduces the S/N ratio compared to the mean quiet Sun and hence hampers the retrieval of $B$.
%

\begin{figure*}
\centering
\includegraphics[width=0.79\linewidth]{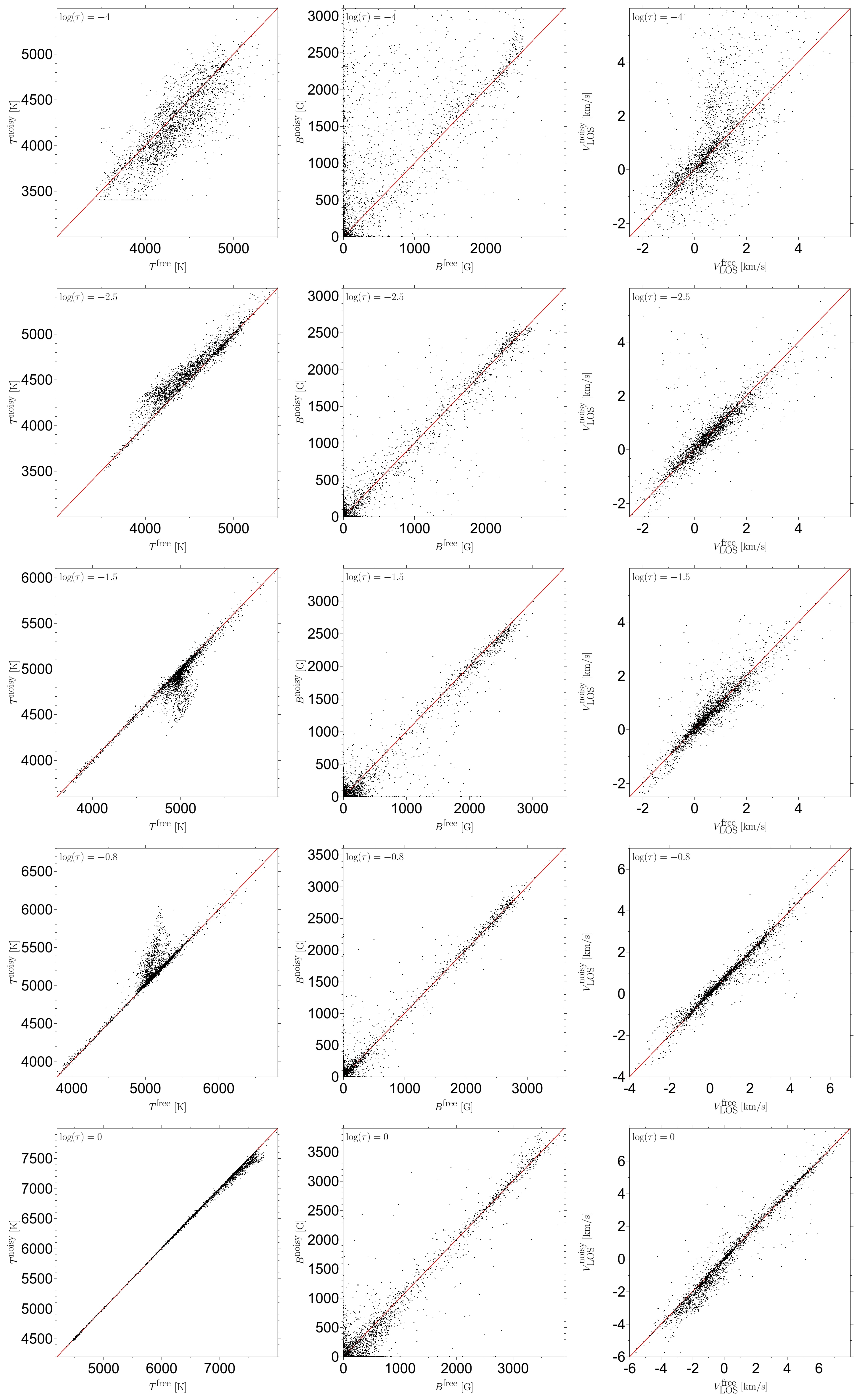}
\caption{Same as Fig.~\ref{Fig6}, but for the spectral region $6300.8\,\rm{\AA} - 6303.2\,\rm{\AA}$ that only contains the Fe\,{\sc i}
6301.5\,\AA{} and 6302.5\,\AA{} line pair.}
\label{Fig7}
\end{figure*}

In our second example we display the scatterplots of the frequently used double-line approach for the spectral region around 6302\,\AA{} (Fig.~\ref{Fig7}).
Again, we choose the inversion with five optical depth nodes for a direct comparison with Fig.~\ref{Fig6}. The $T$ and $v_{\mathrm LOS}$ values display a
considerably larger scatter than at 3140\,\AA{} (Fig.~\ref{Fig6}). Only the scatter in $B$ is approximately equally large, at least for the lower four depth nodes.
In the red, the $B$ scatter is not concentrated at strong fields, but is uniformly distributed over the entire $B$ range because the intensity in the pore drops
only to 17\,\% (due to the smaller temperature sensitivity of the Planck function in the red part of the spectrum), which leads to a better S/N ratio in the red
than in the NUV. In the upper photosphere, at $\log\tau=-4$, the scatter of all quantities is by far larger in the red than in the NUV.
%

\subsection{Synthetic spectra versus observed spectra}

\begin{figure*}
\centering
\includegraphics[width=\linewidth]{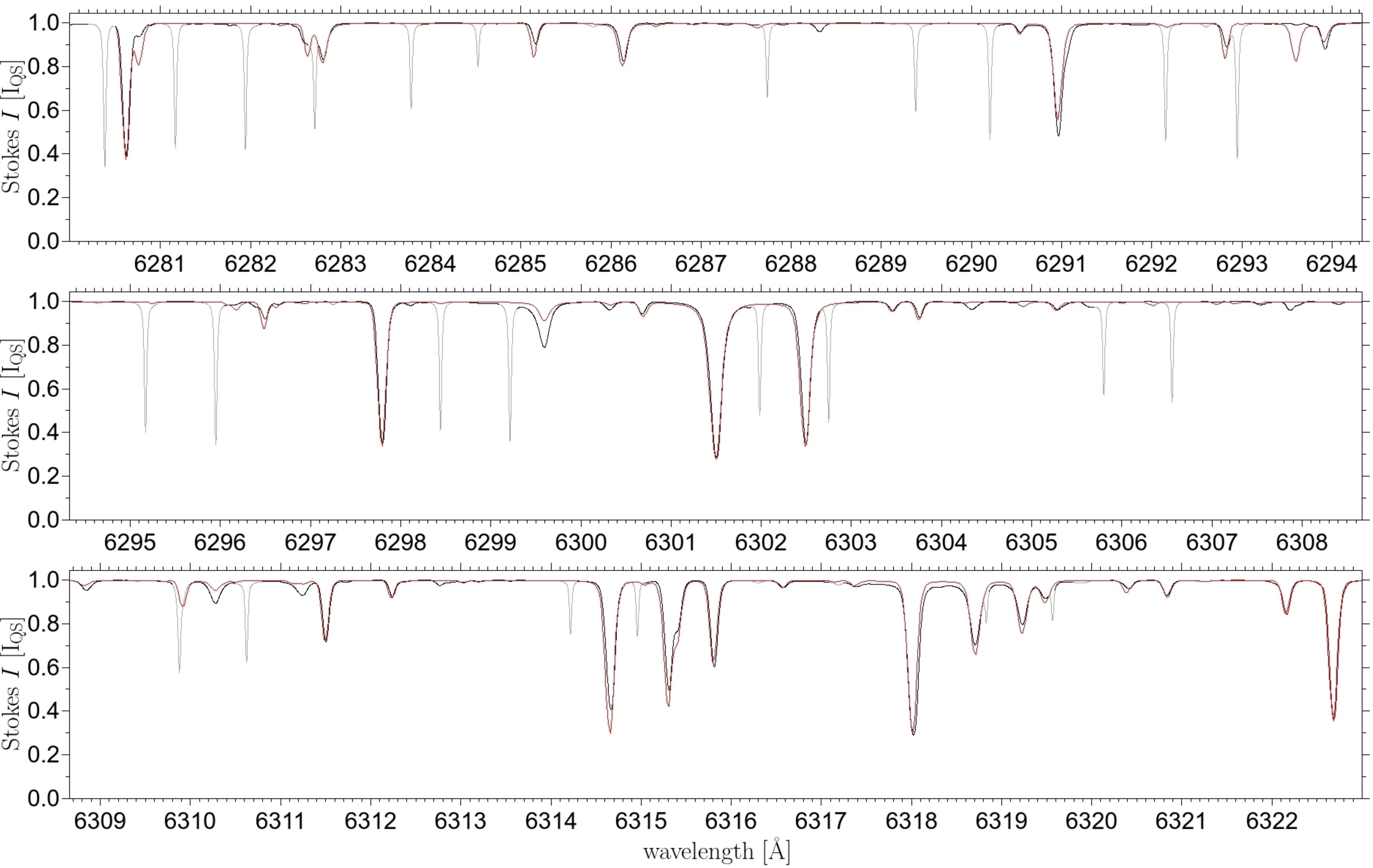}
\caption{Excerpts from the Jungfraujoch atlas (solar lines in black color, telluric lines in gray color) and from spatially averaged synthetic spectra
(red lines; see main text for details).}
\label{Fig8}
\end{figure*}

\begin{figure*}
\centering
\includegraphics[width=\linewidth]{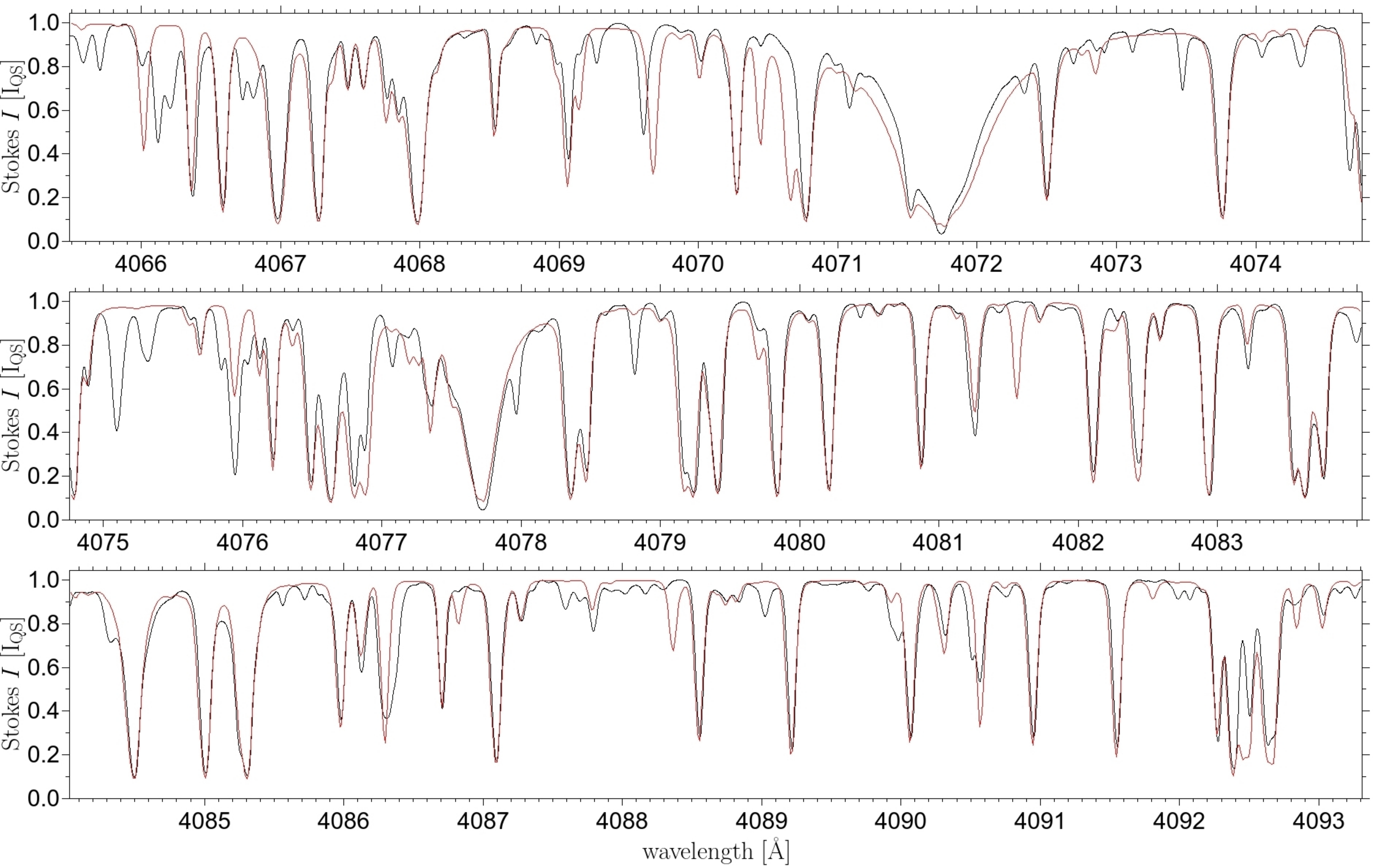}
\caption{Same as Fig.~\ref{Fig8}, but for the spectral region around 4080\,\AA{}.}
\label{Fig9}
\end{figure*}

\begin{figure*}
\centering
\includegraphics[width=\linewidth]{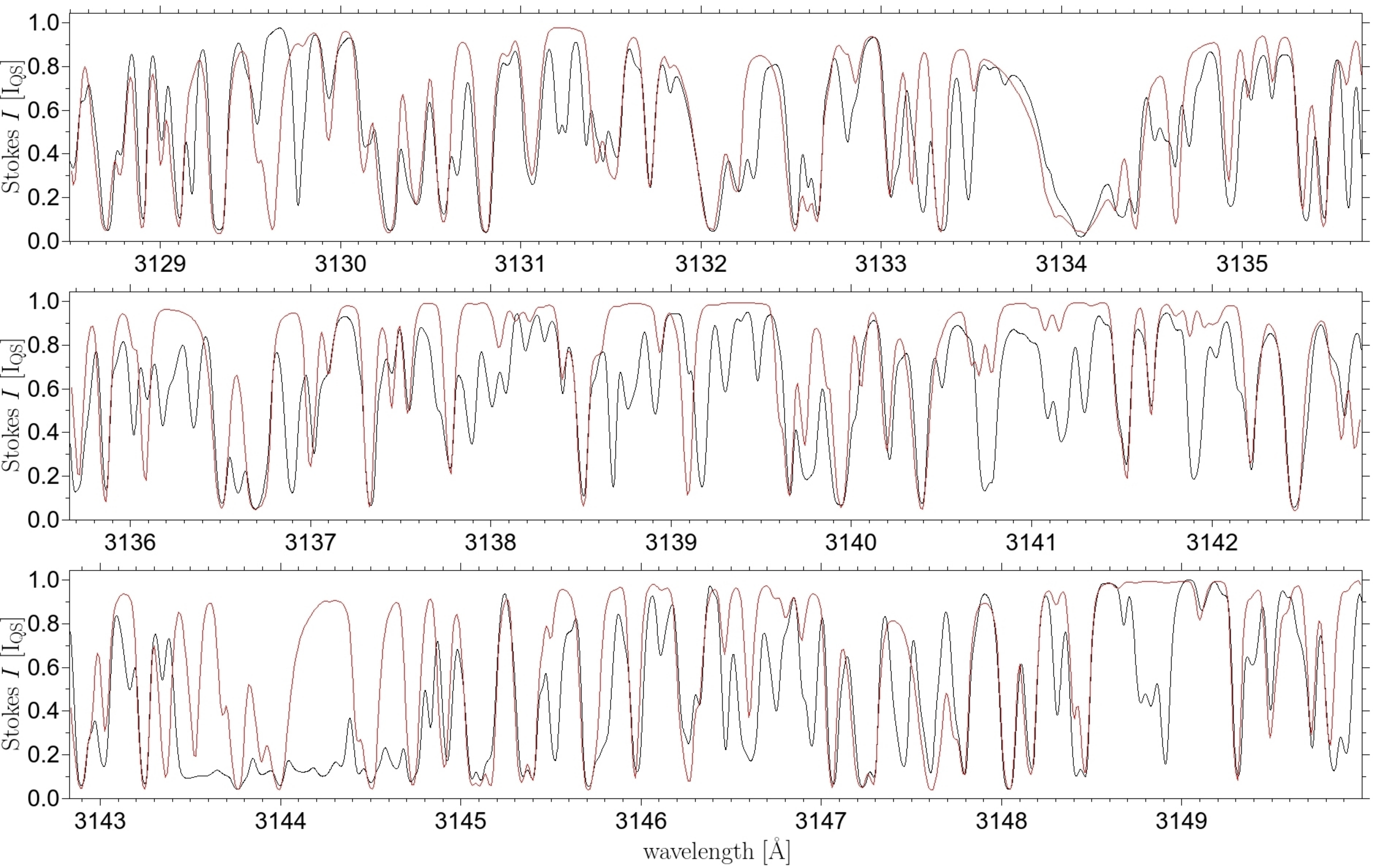}
\caption{Same as Fig.~\ref{Fig8}, but for the spectral region around 3140\,\AA{}.}
\label{Fig10}
\end{figure*}

For a practical application of the many-line approach, not only the trade-off between the photon noise and the number of spectral lines is of importance,
but also the following questions play a role: How complete is the list of atomic and molecular spectral lines used for the synthesis and how good is our knowledge of the line parameters
(e.g., the oscillator strengths)? In Figs.~\ref{Fig8}~to~\ref{Fig10} we contrast synthesized spectra with the Jungfraujoch atlas \citep{Delbouille1973}
in order to gain an impression of the completeness and reliability of our syntheses of the solar spectrum. The Jungfraujoch atlas is a spatially averaged spectrum recorded
at DC with a spectral resolving power of 216000 \citep{Doerr2016}. To imitate this, we use a non-gray MURaM simulation box consisting of
288\,$\times$\,288\,$\times$\,100 cells each of the size 20\,km\,$\times$\,20\,km\,$\times$\,16\,km with a mean vertical magnetic field strength of 100\,G,
and calculate a synthetic spectrum for each of the 288\,$\times$\,288 atmospheres. After a spectral degradation corresponding to a resolving power of 216000, a spatial
averaging of all spectra provides the red lines in Figs.~\ref{Fig8}~to~\ref{Fig10}. The fact that a simulation with a relatively strong mean field provides a good fit,
suggests that the Jungfraujoch atlas sampled enhanced network, in addition to quiet Sun (at least at some wavelengths).

Fig.~\ref{Fig8} compares the Jungfraujoch atlas with the mean synthetic Stokes~$I$ profile for the wavelength region $6280\,\rm{\AA} - 6323\,\rm{\AA}$.
Many spectral lines can be reproduced quite well, e.g. the Fe\,{\sc i} lines at 6280.616\,\AA{}, 6286.132\,\AA{}, 6297.792\,\AA{}, and 6322.687\,\AA{},
or the Ti\,{\sc i} lines at 6303.757\,\AA{} and 6312.237\,\AA{}. Possibly the atomic parameters of some lines are not accurately known, e.g. for the Si\,{\sc i}
line at 6299.599\,\AA{}. A systematic deviation of synthetic lines of a particular element from the observed solar spectrum (see, e.g., the V\,{\sc i}
lines at 6285.150\,\AA{}, 6292.825\,\AA{}, and 6296.487\,\AA{}) can also indicate that the employed solar abundance of the element is inaccurate. Abundances
and line parameters such as the oscillator strength can be fitted with SPINOR, but this was not done in this study. Here we always used the abundances of
\citet{Grevesse1998} and the line parameters of the VALD and Kurucz databases. If a spectral line is contained in both databases and the databases provide
different values for a particular line parameter then we simply used the mean value for our synthesis. Finally, the solar spectrum can contain spectral lines
that are not yet identified or whose atomic data are at least not present in the employed databases, see e.g., the line at 6304.3\,\AA{}.

For the wavelength region $4065.49\,\rm{\AA} - 4093.31\,\rm{\AA}$ the comparison between the observed and synthesized spectrum is displayed in Fig.~\ref{Fig9}.
A large fraction of the spectrum is well reproduced by our synthesis, in particular the Mn\,{\sc i} 4070.279\,\AA{} and the Fe\,{\sc i} 4080.875\,\AA{} line
with their strong polarization signals (see Fig.~\ref{Fig4}) show a good match. Nonetheless, the number of unidentified lines, or lines whose parameters are
only poorly known increases with decreasing wavelength. The Fe\,{\sc i} 4071.737\,\AA{} line exhibits a further type of mismatch. The wings of
the synthetic line are too wide on both sides. This can possibly due to uncertainties in the damping constants of the spectral lines that are used by SPINOR
and taken from \citet{Anstee1995,Barklem1997,Barklem1998}. However, other sources of such a discrepancy cannot be ruled out.

Finally, Fig.~\ref{Fig10} contrasts the spectra in the wavelength region $3128.5\,\rm{\AA} - 3150\,\rm{\AA}$. The discrepancy between the synthetic and the observed
spectrum is far more extreme than for the two other spectral ranges. With our current knowledge, a Stokes inversion makes sense in this spectral range only if it is
limited to sub-intervals showing a good match, so that the many-line approach cannot yet be fully exploited. It is, however, important that an instrument
like SUSI records for the first time spatially resolved full-Stokes spectra, which can help improve our knowledge of spectral lines in this
region. \citet{Stenflo1984} have shown how Stokes spectra can be used to test the identification of lines and find hidden blends more sensitively than
by using Stokes~$I$ alone. \citet{Solanki1985} also showed how Stokes spectra can be used to determine Land\'e factors of lines, which constrains their identification.

\section{Summary and conclusions}\label{Summary}

We synthesized the atmospheres of a state-of-the-art 3D MHD simulation of typical solar surface features (granulation, intergranular lanes, bright points, pores)
in the red spectral range and added photon noise at a level expected for a modern slit spectropolarimeter. The noisy Stokes profiles were inverted via the SPINOR
inversion code and the resulting atmospheric quantities were compared with the inversion results of the noise-free profiles. Synthesis and inversion were
carried out for two spectral regions of a different width. Firstly, we followed the double-line approach for the wavelength range $6300.8\,\rm{\AA} - 6303.2\,\rm{\AA}$,
which only contains the traditional Fe\,{\sc i} line pair. Secondly, we took the many-line approach for the range $6280\,\rm{\AA} - 6323\,\rm{\AA}$, which can be
recorded with a 2K detector at a spectral resolving power of 150000 and contains not only the Fe\,{\sc i} 6301.5\,\AA{} and 6302.5\,\AA{} line pair, but
also 108 further spectral lines identified with the help of the VALD and Kurucz databases.

Our results show the many-line approach to provide clearly superior inversion results over the traditional approach of observing only a few spectral lines.
The loss of information due to the unavoidable photon noise can be compensated to a certain degree by the consideration of many spectral lines. 
With the help of the many-line approach, $T$, $B$, and $v_{\mathrm LOS}$ can be determined much more accurately. For particular optical depth nodes, the
improvement is up to a factor of two for $B$, a factor of four for $v_{\mathrm LOS}$, and a factor of seven for $T$. We conclude that the development of
a future spectropolarimeter should always consider the possibility of recording as many spectral lines as possible.

Encouraged by the success in the red spectral range, we then asked the question, if the worse S/N ratio in the violet and NUV spectral range due to the shape of the
Planck function can be compensated by the increased number of spectral lines we find in this wavelength domain. To this end, we synthesized all atomic spectral
lines present in the VALD and Kurucz databases in the spectral range $3000\,\rm{\AA} - 4300\,\rm{\AA}$, together with some known molecular lines, for a standard
atmosphere with a magnetic field. We looked for auspicious wavelength regions that contain as many spectral lines with strong polarization signals as possible
and can be recorded with a 2K $\times$ 2K pixels detector. We selected two regions out of several candidates: the NUV region $3128.5\,\rm{\AA} - 3150\,\rm{\AA}$ with 371
identified spectral lines and in the violet the region $4065.49\,\rm{\AA} - 4093.31\,\rm{\AA}$ with 328 lines. We then treated the two spectral regions
in the same manner as in the red, i.e. we synthesized the atmospheres of the 3D MHD simulation, contaminated the Stokes profiles with photon noise, and
inverted them via the many-line approach.

We note that in this study only the influence of the photon noise on the quality of the inversion results is investigated. Other effects such as
stray light, wavefront aberrations, polarimetric efficiency of the grating, missing or imperfect line parameters, or deviations in the line formation
from LTE are ignored. Our inversions included the Zeeman effect, but not the Hanle effect. Also, a possible absorption of the NUV wavelengths in the residual
terrestrial atmosphere at float altitude of a stratospheric balloon is neglected in this study. In particular a possible stray-light contamination of
spectropolarimetric data can influence the quality of the inversion results \citep{Riethmueller2014,Riethmueller2017b}, but is difficult to assess because
for observations taken from above the terrestrial atmosphere the stray-light PSF depends only on the properties of the telescope and instrument. For the
third \sunrise{} flight we therefore intend to use an additional calibration target that was designed to measure the stray-light properties of the
instrumentation.

A comparison of the many-line approach for the three wavelength regions under study (one in the NUV, one in the violet, and one in the red) shows that
in the upper photosphere, at $\log\tau=-4$, all atmospheric parameters can be determined considerably more accurate in the NUV and violet than in the red
spectral region. Consequently, spectropolarimetry at short wavelengths allows increasing the height coverage of measurements of the
solar atmospheric quantities. Lower down, at $\log\tau=-2.5, -1.5, -0.8, 0$, we find a higher accuracy for the determination of $T$ for the violet and UV wavelength bands.
For $B$ and $v_{\mathrm LOS}$ we see a differentiated picture. While in the violet the accuracy of the retrieved $B$ and $v_{\mathrm LOS}$ is almost equally good as in the red,
in the NUV we obtain $v_{\mathrm LOS}$ more accurately, but at the price of a slightly lower accuracy for $B$. The optimum is hence reached for the region around
4080\,\AA{} where all considered atmospheric quantities could be determined accurately. This shows that the Zeeman splitting (which scales with the square of
the wavelength) is not the only important quantity in the assessment of the inversion quality, but also the height range over which lines are formed and the amplitude
of the polarization signals, which is stronger in the 4080\,\AA{} region than around 6302\,\AA{}.

We averaged our synthetic Stokes~$I$ spectra over a quiet-Sun region and contrasted the mean spectrum with the observations of the Jungfraujoch atlas. We found a
relatively good match for the $4080\,\rm{\AA}$ wavelength band but a significant discrepancy for the band at $3140\,\rm{\AA}$. This demonstrates that many spectral
lines in the NUV are either not identified or the line parameters are not well known. We encourage the solar and atomic physics community to improve our knowledge about
spectral lines in the wavelength range between $3000\,\rm{\AA} - 4000\,\rm{\AA}$, e.g., by laboratory measurements, so that Stokes inversions of future observations
in the NUV can benefit from.

We conclude from the current study that spectropolarimetry makes sense even at short wavelengths in the violet and NUV spectral range, if the observation is not
limited to only a few spectral lines but dozens, or better still hundreds of lines are measured simultaneously. We found a particularly auspicious spectral
region around 4080\,\AA{} that contains a whole host of strongly polarized lines. This allows the atmospheric parameters to be determined more accurately than in
the red at 6302\,\AA{}. In principle, the Sun can be observed at 4080\,\AA{} from the ground, but this is hampered by the fact that most ground-based telescopes
show a low transmission in the violet, and also by the strong scattering of violet light in the terrestrial atmosphere. By developing a spectropolarimeter
optimized for the range $3000\,\rm{\AA} - 4300\,\rm{\AA}$ and using it during the planned third flight of the balloon-borne observatory \sunrise{}, these
disadvantages can be avoided.

\begin{acknowledgements}
We would like to thank M. van Noort and A. Feller for their valuable contribution to the photon budget calculation and for their help
in setting up SPINOR for the many-line approach. We also thank Tiago Pereira for his precious comments and suggestions to improve the paper.
This work has made use of the VALD database, operated at Uppsala University, the Institute of Astronomy RAS in Moscow, and the University of Vienna.
This project has received funding from the European Research Council (ERC) under the European Union's Horizon 2020 research and innovation program
(grant agreement No. 695075) and has been supported by the BK21 plus program through the National Research Foundation (NRF) funded by the Ministry
of Education of Korea.
\end{acknowledgements}


\end{document}